\documentclass[9pt,twocolumn,twoside]{osajnl}
\journal{ol}
\setboolean{shortarticle}{true}
\usepackage{graphicx}
\usepackage{amssymb}
\usepackage{amsmath}
\usepackage[english]{babel}

\title{Retrieving nonlinear refractive index of nanocomposites using finite-difference time-domain simulations}
\author[1,*]{Andrey V. Panov}
\affil[1]{Institute of Automation and Control Processes,
Far East Branch of Russian Academy of Sciences,
5, Radio st., Vladivostok, 690041, Russian Federation}
\affil[*]{Corresponding author: panov@iacp.dvo.ru}
\dates{Received 12 April 2018; accepted 19 April 2018; published 18 May 2018}
\ociscodes{(190.4400)   Nonlinear optics, materials; (260.2065)   Effective medium theory; (160.4330)   Nonlinear optical materials}
\setboolean{displaycopyright}{true}
\renewcommand*{\journalname}{Vol. 43, No. 11 (2018) // Optics Letters}
\doi{\url{https://doi.org/10.1364/OL.43.002515}}

\begin{document}

\begin{abstract}
 
In this Letter, it is proposed a method which utilizes three-dimensional finite-difference time-domain (FDTD) simulations of light propagation for restoring the effective Kerr nonlinearity of nanocomposite media. In this approach, a dependence of the phase shift of the transmitted light on the input irradiance is exploited. 
The reconstructed values of the real parts of the nonlinear refractive index of a structure of randomly arranged spheres are in good agreement with the predictions of the effective medium approximations.
\end{abstract}

\maketitle

In recent decades, considerable attention has been given to the study of the composite materials with nonlinear optical properties. Particularly, metamaterials with the tailored nonlinear optical response are promising materials for a plethora of applications, e.g. optical switching, super-resolution imaging and transformation optics.
In order to accelerate the development of the nonlinear optical composites, their properties should be modeled theoretically.

In theory, the optical properties of nanocomposites are usually treated with the effective medium approximations which replace the material containing subwavelength inclusions by homogenized one.
There exist effective medium theories describing the nonlinear optical characteristics of the nanocomposites with several geometries: ellipsoidal or spherical inclusions in host \cite{Stroud88,Agarwal88,Rukhlenko12}, layered structures \cite{Boyd96}. It is hard to obtain analytical expressions for more complicated materials. Alternatively, the nonlinear optical properties of such nanocomposites can be represented numerically. 
Meng et al. \cite{Meng08} used two-dimensional FDTD technique for simulating Z-scan experiments and investigating the nonlinearity enhancement in one-dimensional photonic crystals. Del Hoyo et al. \cite{Hoyo15} proposed a method based on the simple monitoring of the nonlinear beam shaping against numerical solutions of the scalar nonlinear Schr\"odinger equation.
Their method provides a way of estimating the effective and the nonlinear absorption coefficient in homogeneous dielectrics with ultra-short laser pulses.
Liu and Song \cite{Liu13} exploited the theory describing the spectrum of a light pulse in filament \cite{Cubeddu70} for retrieving the nonlinear refractive index of the isotropic homogeneous material with the FDTD simulations.
Nowadays, for experimental measuring the strength of the Kerr nonlinearity of an optical material the z-scan measurement technique employing the light phase change due to the nonlinear refraction and absorption is usually used \cite{Sheik-Bahae90}. In this work, it is shown that the dependence of the light phase change on the input irradiance power of the Gaussian beam can be used for estimating the real part of the nonlinear refraction coefficient of inhomogeneous medium with FDTD modeling.

\begin{figure}
{\centering\includegraphics[scale=0.55]{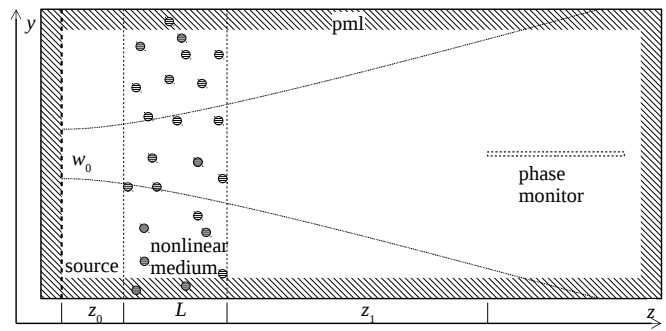}\par} 
\caption{\label{nlphasefig} Schematic of the FDTD simulation. The Gaussian beam propagates from the source through a nonlinear specimen along the $z$ direction. Later the data accumulated at several points in the phase monitor allows one to calculate the phase shift introduced by the specimen. The phase shift carries the information about the nonlinear refractive index of the sample. The computational domain is bounded by the perfect matched layers.}
\end{figure} 

In these simulations, the laser Gaussian beam propagating along the $z$-axis illuminates a thin sample with the optical nonlinearity (see Fig.~\ref{nlphasefig}). The electric field of the Gaussian beam having a waist at $z=0$ is given by:
\begin{multline}
\label{Egaussianbeam}
E(r,z)=E_0 \frac{w_0}{w(z)}\exp\left\{-\frac{r^2}{w(z)^2} \right\}    \times \\
\exp\left\{-i\left\{kz +  \frac{kr^2}{2z\left[ 1+\left( z_R/ z\right)^2 \right]} - \arctan\left( \frac{z}{z_R}\right)\right\} \right\}
\end{multline}
with the electric field amplitude $E_0$ and beam radius $w_0$ at the beam waist, the wavenumber $k=2\pi/\lambda$, the Rayleigh length $z_R=\pi w_0^2/\lambda$ and    radial distance from the center axis of the beam $r$. 
The light intensity distribution is written as
\begin{equation}
 I(r,z)=I_0 \left(\frac{w_0}{w(z)}\right)^2 \exp \left(- \frac{2r^2}{w(z)^2} \right) ,
\end{equation} 
$I_0\propto |E_0|^2$ is the intensity at the center of the beam at its waist.
The phase shift of the Gaussian beam at the beam axis is 
\begin{equation}\label{phigaussianbeam}
 \phi=kz - \arctan\left( \frac{z}{z_R}\right).
\end{equation} 
The term with the $\arctan$ function describes the Gouy phase shift.

The laser beam passes through a  sample with  the Kerr nonlinearity having the refractive index,
\begin{equation}
 n=n_0+n_2 I,
\end{equation} 
where $n_0$ is the linear refractive index, $n_2$ is the second-order nonlinear refractive index, and $I$ is the intensity of the wave. The second term is assumed to be much less than $n_0$.
The Kerr nonlinearity in the sample adds the intensity dependent phase change at the beam axis
\begin{equation}
 \Delta\phi_{(3)} = k n_2 \int_{z_0}^{z_0+L} I(x',y,'z')dx'dy'dz' = k n_2 J \propto I_0.
\end{equation} 
Integral $J$ is taken near the beam axis within the nonlinear material and can be evaluated numerically during computations. It should be noted that the FDTD calculations are performed in dimensionless units. So the computed $\Delta\phi_{(3)}$ must be compared with the nonlinear phase shift of the reference specimen with the known $n_2$.

The phase change on the beam axis transmitted through a dielectric slab for fundamental Gaussian mode is \cite{Antar74,Ooya75}
\begin{equation}
 \label{deltaphi1_n0}
 \Delta\phi_{(1)}=kL(n_0-n_b)+\arctan\frac{z}{z_R}-\arctan\frac{z+L(n_b/n_0-1)}{z_R},
\end{equation} 
here $n_b$ is the refractive index of the ambient medium.

The net phase change $\Delta\phi(I_0)=\Delta\phi_{(1)}+\Delta\phi_{(3)}(I_0)$ carries information about the refractive index of the nonlinear sample. Thus, by knowing $\Delta\phi(I_0)$, it is possible to estimate $n_0$ and $n_2$ of a modeled medium. Further, the feasibility of this technique will be demonstrated.

\begin{figure}[t]
{\centering\includegraphics[scale=1]{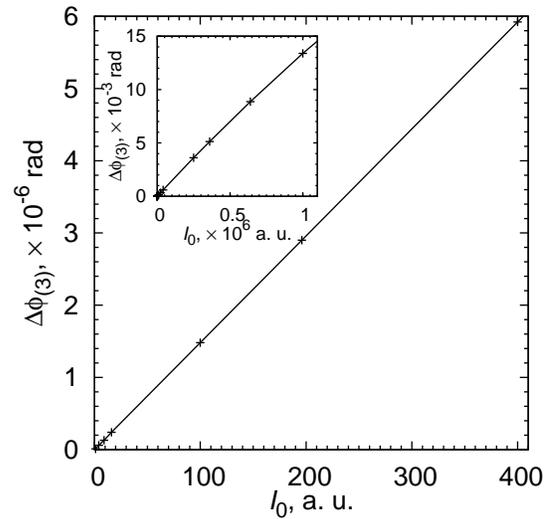}\par} 
\caption{\label{phasevsintesityfig}Dependence of the phase change due to the Kerr nonlinearity $\Delta\phi_{(3)}$ on the input intensity at the specific point in the phase monitor. The studied specimen consists of the spheres with the radius of $20$~nm, $n_0=1.5$, and $\chi^{(3)}=10^{-11}$~esu, which are randomly arranged in air with the volume fraction of 0.0654. The point in the phase monitor was located at the distance of 7.7~$\mu$m from the source. The intensity is expressed in arbitrary units. The linear phase shift $\Delta\phi_{(1)}$ introduced by the medium was $0.228318$~rad and it does not depend on the intensity. As one can see, at small intensities, $\Delta\phi_{(3)}$ varies linearly with $I_0$.}
\end{figure}

The FDTD simulation of the phase change introduced by the nonlinear sample is done as follows. 
A continuous planar light source with the Gaussian profile of the electric current excites the beam propagating along the $z$ direction in an ambient medium (e.g. vacuum). 
Further, the beam falls perpendicularly on the specimen. Then, the transmitted beam again propagates through the surrounding medium. It is assumed that the studied sample does not significantly distort the beam. 
A phase monitor is located along beam axis at reasonably large distance $z_1$ from the sample so that $z_1 \gg z_R$. Here at several points, the instant electric field component in the $y$ direction $E_y(t)$ is accumulated for some time $\Delta t$. After the FDTD simulation, the data is processed with the discrete Fourier transform, the complex argument of this transform at the frequency of interest represents the phase of the wave at a fixed point of the phase monitor. The simulations are run twice, once with the studied structure, and once for the ambient medium only. 
The difference between two simulations for  a certain point gives the calculated $\Delta\phi$. 
By varying $I_0$, it is possible to estimate $n_2$ of the sample. 
Fig.~\ref{phasevsintesityfig} depicts the dependence of $\Delta\phi_{(3)}$ on intensity for  a certain point of the phase monitor. 
For low intensities of the beam, this dependence is linear. 
The linear fit of the calculated values of $\Delta\phi(I_0)$ gives $\Delta\phi_{(1)}$ and $\Delta\phi_{(3)}(I_0)$. 
By comparing the slope of $\Delta\phi_{(3)}(I_0)$ with one of the reference specimen, it is possible to calculate $n_2$ at the point. 
After solving
Eq.~(\ref{deltaphi1_n0}), the value of $n_0$ is available. The value of the integral $J$ can be computed using the sample with known $n_2$. The several magnitudes of $n_2$ at different points of the phase monitor make it possible to estimate the mean value and the standard deviation of the non-linear refractive index.

For simulations, the Massachusetts Institute of Technology (MIT) Electromagnetic Equation Propagation (MEEP) \cite{OskooiRo10} FDTD solver was utilized. The size of the computational domain was typically $1.6\times 1.6\times 8$~$\mu$m, the boundaries around the domain  were handled by perfectly matched layers (PML). The space resolution for the FDTD modeling was 2.5~nm, tests with other resolutions demonstrated similar results. The size of the specimen along the $z$ direction $L$ was $0.8$~$\mu$m, its other dimensions were limited with the computational domain. The distance between the sample and phase monitor was $6.8$~$\mu$m. The wavelength of the simulated light was 667~nm, the beam radius $w_0$ at the beam waist was 283 or 600~nm. 
The first beam size was not quite paraxial ($kw_0/\sqrt{2}=1.9$, see \cite{Nemoto90}) but the phase dependence on $z$ at the beam axis at large $z$ agrees well with one of the paraxial Gaussian beam \cite{Nemoto90}. 
The Gaussian beam with $w_0=600$~nm ($kw_0/\sqrt{2}=4$) was shown to be paraxial \cite{Nemoto90}. 
The FDTD simulations of the $1.6\times 1.6\times 8$~$\mu$m computational domain were performed on supercomputers with at least 256 GB of RAM, the size of RAM required for modeling $3.2\times 3.2\times 8$~$\mu$m domain is four times larger. 
After the start of the continuous wave source, simulations are performed for three periods of the wave in order to reach a steady state. 
Then, there is a need for modeling until the light wave propagates through whole computational domain. 
Subsequently, the data for the Fourier transform is accumulated for nine periods of the wave. 
This time interval enables one to obtain the Fourier transform of $E_y$ at the same frequency as emitted by the source. 
Further, the saved data of $E_y$ is processed separately.

\begin{table}
\renewcommand{\arraystretch}{1}
{\centering 
\begin{tabular}{llp{3.5em}p{7em}lll}
\hline
$n_0$	&$n_{0\,\mathrm{rec}}$&$\chi^{(3)}$, $10^{-11}$~esu&$\chi^{(3)}_{\mathrm{rec}}$, $10^{-11}$~esu	&$w_0$, nm&$n_{0\,\mathrm{ref}}$\\
\hline
1.05	&1.0501	&1	&$0.98	\pm 0.04$	&283&	1.1\\
1.2	&1.2000	&1	&$1.03	\pm 0.04$	&283&	1.1\\
1.2	&1.2000	&$-1$	&$-1.03	\pm 0.04$	&283&	1.1\\
1.5	&1.4989	&1	&$0.94	\pm 0.04$	&283&	1.1\\
1.5	&1.4989	&0.1	&$0.094	\pm 0.004$	&283&	1.1\\
2	&1.9931	&1	&$0.96	\pm 0.06$	&283&	1.5\\
2.2	&2.2008	&1	&$1.02	\pm 0.07$	&283&	2.1\\
1.2	&1.1994	&1	&$0.96	\pm 0.05$	&600&	1.1\\
1.05	&1.0500	&1	&$1.00	\pm 0.05$	&600&	1.1\\
1.03	&1.0301	&0.018	&$0.0172	\pm 0.0005$	&283&	1.1\\
1.059	&1.0592	&0.041	&$0.040	\pm 0.001$	&283&	1.1\\
\hline

\end{tabular}

\par}
\caption{\label{resnhomogeneoustab}The results of the reconstruction of linear refractive index $n_0$, nonlinear susceptibility $\chi^{(3)}$ and its standard deviation for the homogeneous specimens. The nonlinear susceptibility is linearly related to the second-order nonlinear refractive index $n_2$. The first and the third columns display the true values of $n_0$ and $\chi^{(3)}$, the second and the fourth columns exhibit the retrieved magnitudes. Here $n_{0\,\mathrm{ref}}$ is the refractive index of the reference specimen. The ambient medium with $n_b=2$ was used for the FDTD simulations of transmission through the layer with $n_0>2$.}
\end{table}


First, to test the described technique, the optical nonlinearity of the continuous sample should be modeled.
Table~\ref{resnhomogeneoustab} presents the results of the reconstruction of $n_0$ and nonlinear susceptibility $\chi^{(3)}=n_2 n_0 /(4 \pi)$ for the homogeneous samples. The reference specimens for the computations have $\chi^{(3)}=10^{-11}$~esu. As is clearly seen from Table~\ref{resnhomogeneoustab}, the  values of $n_0$ and $\chi^{(3)}$ are reconstructed reasonably well. The error of the restoration for the specimens with $n_0=1.5$ and $n_0=2$ is larger since the high refractive index contrast with the ambient medium is responsible for exciting higher-order modes of the transmitted beam \cite{Ooya75}. Hence, $\Delta\phi_{(1)}$ measurably differs from one defined by Eq.~(\ref{deltaphi1_n0}). At large distances $\Delta\phi_{(1)}$ tends to $kL(n_0-n_b)$ even for the higher-order modes. Thus, the enlarging the computational domain may reduce the error in this case. 
Otherwise,  the surrounding medium at $z<z_0$ and $z>z_0+L$ with $n_b$ close to $n_0$ may be utilized. 
For example, the layers with $n_0>2$ were modeled with the background medium having $n_b=2$ and $\chi^{(3)}$ was restored well. 
It is to be noted that the composites even with the high index inclusions typically have the effective index of refraction below 2 owing to their moderate volume fractions.
Also the values for $w_0=600$~nm and $1.6\times 1.6\times 8$~$\mu$m size of the computational domain are retrieved worse as the intensity at its boundaries is larger than for $w_0=283$~nm. The fields anyway are distorted by the boundaries of the domain.

\begin{table*}[tb]
{
\renewcommand{\arraystretch}{1}
\centering 
\begin{tabular}{llllp{2.3cm}p{1.8cm}p{1.8cm}p{1.8cm}p{1.8cm}}
\hline
$f$	&$n_{0\,\mathrm{rec}}$	&$n_{0\,\mathrm{MG}}$	&$n_{0\,\mathrm{B}}$	&$\chi^{(3)}_{\mathrm{rec}}$, $10^{-13}$~esu	&$\chi^{(3)}_{\mathrm{SH}}$, $10^{-13}$~esu	&$\chi^{(3)}_{\mathrm{AG}}$, $10^{-13}$~esu	&$\chi^{(3)}_{\mathrm{RZPA}}$, $10^{-13}$~esu	&$\chi^{(3)}_{\mathrm{in}}$, $10^{-11}$~esu	\\
\hline
\multicolumn{9}{l}{\hspace{3ex}$w_0=283$~nm}\\
0.0164	&1.0072	&1.0072	&1.0072	&$0.39	\pm 0.01$	&0.41	&0.41	&0.42	&1	\\
0.0164	&1.0073	&1.0072	&1.0072	&$0.44	\pm 0.01$	&0.41	&0.41	&0.42	&1	\\
0.0164	&1.0075	&1.0072	&1.0072	&$0.46	\pm 0.02$	&0.41	&0.41	&0.42	&1	\\
0.0164	&1.0073	&1.0072	&1.0072	&$0.41	\pm 0.01$	&0.41	&0.41	&0.42	&1	\\
0.0327	&1.0151	&1.0145	&1.0146	&$0.85	\pm 0.02$	&0.81	&0.84	&0.87	&1	\\
0.0327	&1.0147	&1.0145	&1.0146	&$0.92	\pm 0.03$	&0.81	&0.84	&0.87	&1	\\
0.0327	&1.0161	&1.0145	&1.0146	&$0.92	\pm 0.03$	&0.81	&0.84	&0.87	&1	\\
0.0327	&1.0148	&1.0145	&1.0146	&$0.92	\pm 0.03$	&0.81	&0.84	&0.87	&1	\\
0.0327	&1.0147	&1.0145	&1.0146	&$0.092	\pm 0.003$	&0.081	&0.084	&0.087	&0.1	\\
0.0654	&1.0303	&1.0290	&1.0293	&$1.81	\pm 0.04$	&1.62	&1.76	&1.87	&1	\\
0.0654	&1.0298	&1.0290	&1.0293	&$-1.82	\pm 0.04$	&$-1.62$	&$-1.76$	&$-1.87$	&$-1$	\\
0.1309	&1.0591	&1.0584	&1.0595	&$4.10	\pm 0.09$	&3.25	&3.80	&4.28	&1	\\
0.2618	&1.1196	&1.1182	&1.1222	&$9.93	\pm 0.22$	&6.50	&8.96	&10.97	&1	\\
0.2618	&1.1196	&1.1182	&1.1222	&$-9.93	\pm 0.22$	&$-6.50$	&$-8.96$	&$-10.97$	&$-1$	\\
0.1306	&1.0564	&1.0582	&1.0594	&$3.88	\pm 0.09$	&3.24	&3.79	&4.27	&1	\\
\multicolumn{9}{l}{\hspace{3ex}$w_0=600$~nm}\\
0.0164	&1.0072	&1.0072	&1.0072	&$0.47	\pm 0.02$	&0.41	&0.41	&0.42	&1	\\
0.0164	&1.0076	&1.0072	&1.0072	&$0.48	\pm 0.02$	&0.41	&0.41	&0.42	&1	\\
0.0327	&1.0142	&1.0145	&1.0146	&$0.93	\pm 0.04$	&0.81	&0.84	&0.87	&1	\\
0.0327	&1.0146	&1.0145	&1.0146	&$0.95	\pm 0.04$	&0.81	&0.84	&0.87	&1	\\
0.0654	&1.0293	&1.0290	&1.0293	&$1.91	\pm 0.08$	&1.62	&1.76	&1.87	&1	\\
0.0654	&1.0291	&1.0290	&1.0293	&$-1.92	\pm 0.08$	&$-1.62$	&$-1.76$	&$-1.87$	&$-1$	\\
0.0654	&1.0497	&1.0495	&1.0511	&$0.58	\pm 0.02$	&0.41	&0.47	&0.57	&1	\\
0.1309	&1.0576	&1.0584	&1.0595	&$4.13	\pm 0.18$	&3.25	&3.80	&4.28	&1	\\
0.2618	&1.1152	&1.1182	&1.1222	&$9.65	\pm 0.42$	&6.50	&8.96	&10.97	&1	\\
0.2618	&1.1152	&1.1182	&1.1222	&$-9.65	\pm 0.42$	&$-6.50$	&$-8.96$	&$-10.97$	&$-1$	\\
0.1306	&1.0549	&1.0582	&1.0594	&$3.92	\pm 0.17$	&3.24	&3.79	&4.27	&1	\\
0.2618	&1.1195	&1.1182	&1.1222	&$9.73	\pm 0.17$	&6.50	&8.96	&10.97	&1	\\
\hline

\end{tabular}

\par}
\caption{\label{resnheterogeneoustab}
The linear refractive indexes and nonlinear susceptibilities $\chi^{(3)}$ of mixtures retrieved with the modeling and predicted by the effective medium theories. Here $f$ is the volume fraction (concentration) of inclusions, $n_{0\,\mathrm{rec}}$ is the reconstructed linear refractive index, $n_{0\,\mathrm{MG}}$ and $n_{0\,\mathrm{B}}$ are ones calculated using the Maxwell Garnett \cite{Garnett385} or Bruggeman \cite{Bruggeman35} approximations, $\chi^{(3)}_{\mathrm{rec}}$ is the nonlinear susceptibility retrieved by the FDTD method, for comparison there are presented third-order susceptibilities of mixtures obtained using the nonlinear effective medium approximations: $\chi^{(3)}_{\mathrm{SH}}$\cite{Stroud88}, $\chi^{(3)}_{\mathrm{AG}}$\cite{Agarwal88}, $\chi^{(3)}_{\mathrm{RZPA}}$\cite{Rukhlenko12}; $\chi^{(3)}_{\mathrm{in}}$ is the third-order susceptibility of the inclusions. 
The retrieved values for $f=0.1306$ was calculated for the  aligned cubic inclusions with the edge size of 32~nm. The last row for $w_0=600$~nm was calculated using $3.2\times 3.2\times 8$~$\mu$m size of the computational domain.}
\end{table*}

Evidently, it would be more interesting to study the optical nonlinearity of inhomogeneous samples. The simplest ones are disjoint spheres with equal radii $r_s=20$~nm randomly arranged in space. The inclusions have linear refractive index $n_{0\,\mathrm{in}}=1.5$ and optical Kerr nonlinearity $\chi^{(3)}_{\mathrm{in}}$. The linear refractive index of this system usually is described by the effective medium theory of Maxwell Garnett \cite{Garnett385} or Bruggeman \cite{Bruggeman35}. The retrieved third-order susceptibility of the mixture can be compared with one calculated using effective medium approximations \cite{Stroud88,Agarwal88,Rukhlenko12}. The effective third-order susceptibilities reconstructed with the FDTD simulations and analytically calculated using the effective medium theories are tabulated in Table~\ref{resnheterogeneoustab}.

The results of modeling show that in most cases the retrieved magnitudes of $\chi^{(3)}$ lie in the range between the values predicted by the works \cite{Agarwal88} and \cite{Rukhlenko12}. The model in Ref.~\cite{Stroud88} describes the limit of very diluted nonlinear material. For the low concentrations of the inclusions the discrepancy from the theoretically predicted values of $\chi^{(3)}$ is substantial. This may be associated with fluctuations of inclusion density in the surrounding medium. These fluctuations are more prominent at the lower volume fractions. 
By way of illustration, the results of the simulations for several arrangements of the spheres at low concentrations are presented in Table~\ref{resnheterogeneoustab} as rows with the same volume fraction.
For such mixtures, averaging over many configurations is required. 
For large volume fraction of the inclusions $f=0.2618$ and wide Gaussian beam $w_0=600$~nm, the retrieved linear refractive index significantly differs from the values described by the effective medium theory of Maxwell Garnett \cite{Garnett385} or Bruggeman \cite{Bruggeman35}. This is attributable to the field distortion at the domain boundaries and can be fixed by using the larger computational domain and consequently more computational resources for the simulations. The computed magnitudes of $n_{0}$ and $\chi^{(3)}$ of the aligned cubes ($f=0.1306$) are slightly less than those of the system of spheres ($f=0.1309$) with similar volumes. For comparison, the simulations of two homogeneous slabs with the $n_0$ and $\chi^{(3)}$ reconstructed from the inhomogeneous samples ($f=0.0654$ and $f=0.1309$) are given  in the last rows of Table~\ref{resnhomogeneoustab}. 
They show good agreement with the results extracted for the composites.


As can be seen, the proposed technique is applicable to estimate the optical nonlinearity of composite materials. In contrast, the method of Ref.~\cite{Hoyo15} hardly can be used for this purpose since the shape of the pulse is distorted by the inhomogeneous medium. The theory presented in Ref.~\cite{Cubeddu70} and applied to FDTD modeling in Ref.~\cite{Liu13} was worked out for filaments in liquids, so it will be unlikely suitable for the medium with inclusions.

In summary, the method for retrieving the nonlinear refractive index of composite subwavelength structures is developed. This technique is based on the dependence of the phase shift on the input irradiance in three-dimensional FDTD simulations. The obtained results are shown to be reasonable and correlate with the theoretical values calculated using the effective medium approximations. This method can be applied for studying nonlinear nanocomposites with the shape that cannot be treated analytically.

\subsubsection*{Acknowledgements}
The results were obtained with the use of IACP FEB RAS Shared Resource Center ``Far Eastern Computing Resource'' equipment (https://www.cc.dvo.ru).




\end{document}